\documentclass[aps,prd,twocolumn,nofootinbib,showpacs]{revtex4-2}

\usepackage{graphicx}
\usepackage[colorlinks=true, linkcolor=blue, citecolor=blue, urlcolor=blue]{hyperref}

\begin{document}

\title{Precise determination of the top-quark on-shell mass $M_t$ via its scale-invariant perturbative relation to the top-quark $\overline{\rm MS}$ mass ${\overline m}_t({\overline m}_t)$ }

\author{Xu-Dong Huang$^1$}
\email{huangxd@ihep.ac.cn}

\author{Xing-Gang Wu$^2$}
\email{wuxg@cqu.edu.cn}

\author{Xu-Chang Zheng$^2$}
\email{zhengxc@cqu.edu.cn}

\author{Jiang Yan$^2$}
\email{yjiang@cqu.edu.cn}

\author{Zhi-Fei Wu$^2$}
\email{zfwu@cqu.edu.cn}

\author{Hong-Hao Ma$^3$}
\email{mahonghao@pku.edu.cn}

\affiliation{$^1$ Institute of High Energy Physics, Chinese Academy of Sciences, 19B Yuquan Road, Shijingshan District, Beijing, 100049, P.R. China}
\affiliation{$^2$ Department of Physics, Chongqing Key Laboratory for Strongly Coupled Physics, Chongqing University, Chongqing 401331, P.R. China}
\affiliation{$^3$ Department of Physics, Guangxi Normal University, Guilin 541004, P.R. China}

\date{\today}

\begin{abstract}

It has been shown that the principle of maximum conformality (PMC) provides a systematic way to solve conventional renormalization scheme and scale ambiguities. The scale-fixed predictions for physical observables using the PMC are independent of the choice of renormalization scheme -- a key requirement of renormalization group invariance. In the paper, we derive new degeneracy relations based on the renormalization group equations that involve both the usual $\beta$-function and the quark mass anomalous dimension $\gamma_m$-function, respectively. These new degeneracy relations lead to an improved PMC scale-setting procedures, such that the correct magnitudes of the strong coupling constant and the $\overline{\rm MS}$-running quark mass can be fixed simultaneously. By using the improved PMC scale-setting procedures, the renormalization scale dependence of the $\overline{\rm MS}$-on-shell quark mass relation can be eliminated systematically. Consequently, the top-quark on-shell (or $\overline{\rm MS}$) mass can be determined without conventional renormalization scale ambiguity. Taking the top-quark $\overline{\rm MS}$ mass ${\overline m}_t({\overline m}_t)=162.5^{+2.1}_{-1.5}$ GeV as the input, we obtain $M_t\simeq 172.41^{+2.21}_{-1.57}$ GeV. Here the uncertainties are combined errors with those also from $\Delta \alpha_s(M_Z)$ and the approximate uncertainty stemming from the uncalculated five-loop terms predicted through the Pad\'{e} approximation approach.

\end{abstract}

\maketitle

\section{Introduction}

The top quark is the heaviest elementary particle in the Standard Model (SM), and its mass is one of the most important input parameters of the SM. The largest mass among quarks, or equivalently the strongest Yukawa coupling, implies that the top quark plays a crucial role in governing the stability of the electroweak vacuum. Determining the top-quark mass accurately is helpful for precision test of the SM, for determining whether the vacuum is in stable or meta-stable state, and for the search of new physics beyond the SM. Direct measurements of the top-quark mass have reached a high precision in proton-proton ($pp$) collisions at the LHC~\cite{ATLAS:2016muw, ATLAS:2017lox, ATLAS:2018fwq, CMS:2015lbj, CMS:2017znf, CMS:2018quc, CMS:2018tye, CMS:2021jnp}, which rely on the reconstruction of the top quark decay products and the multipurpose Monte Carlo (MC) event generators. Some other measurements have also been performed in Refs.\cite{CMS:2022emx, ATLAS:2019guf, Sirunyan:2019zvx, ATLAS:2017dhr, ATLAS:2015pfy, CMS:2019fak, CMS:2017mpr, D0:2016txf, D0:2011hwd}. In the theoretical side, there are many attempts to relate the top-quark mass to its on-shell (OS) scheme mass, c.f. Refs.\cite{Buckley:2011ms, Skands:2007zg, Kawabataa:2014osa, Kieseler:2015jzh, Dehnadi:2018hrh, Hoang:2020iah, Khachatryan:2016mqs, Fleming:2007qr}. And it has a $0.5\sim 1$ GeV difference between the top-quark MC mass ($M_t^{\rm MC}$) and the top-quark OS mass ($M_t^{\rm OS}$)~\cite{FerrarioRavasio:2018whr, Butenschoen:2016lpz, ATLAS:2014wva, Moch:2014tta, Juste:2013dsa, Hoang:2014oea, CMS:2018fks}.

The top-quark OS mass has been investigated in detail in Refs.\cite{Hempfling:1994ar, Chetyrkin:1999ys, Jegerlehner:2003py, Martin:2005ch, Galtieri:2011yd, Martin:2016xsp, Nason:2016tiy, Beneke:2016cbu, Bevilacqua:2017ipv, Wang:2017kyd, Garzelli:2023rvx, Cridge:2023ztj, Marquard:2018qqx, Ayala:2019hkn, Ju:2020otc, Wang:2020mel, Cooper-Sarkar:2020twv, Gao:2020nhu, Beneke:2021lkq, Alioli:2022ttk}. The top-quark OS mass can be related to the modified minimal subtraction ($\overline{\rm MS}$) scheme running mass by using the perturbative relation between the top-quark bare mass ($m_{t,0}$) and the renormalized mass in the OS- or $\overline{\rm MS}$- scheme, e.g. $m_{t,0}=Z^{\rm OS}_{m}M_t^{\rm OS}$ and $m_{t,0}=Z^{\overline{\rm MS}}_{m}{\overline m}_t(\mu_r)$. The $Z^{\rm OS}_{m}$ and $Z^{\overline{\rm MS}}_{m}$ are quark mass renormalization constants in the OS- or $\overline{\rm MS}$- scheme, respectively. In perturbative Quantum Chromodynamics (pQCD) theory, the relation between the OS mass and the $\overline{\rm MS}$ mass has been calculated up to four-loop level~\cite{Tarrach:1980up, Gray:1990yh, Chetyrkin:1999qi, Melnikov:2000qh, Jegerlehner:2002em, Jegerlehner:2003sp, Faisst:2004gn, Marquard:2007uj, Marquard:2015qpa, Marquard:2016dcn, Kataev:2018mob, Kataev:2018sjv, Kataev:2018gle}. It allows for determining the OS mass with the help of the experimental result of $\overline{\rm MS}$ mass. During the determination, the key issue is to fix the exact values of the strong coupling constant ($\alpha_s$) and the $\overline{\rm MS}$ mass (e.g. ${\overline m}_q$, $q=c,b,t$ denote the charm, bottom, and top quark, respectively). The scale running behavior of $\alpha_s$ and ${\overline m}_q$ are governed by general renormalization group equations (RGEs) involving both the $\beta$ function and the quark mass anomalous dimension $\gamma_m$, i.e.,
\begin{eqnarray}
\frac{d a_s(\mu_r)} {d\ln\mu^2_r}&=&\beta(a_s)=-\sum_{i=0}^\infty \beta_i a^{i+2}_s(\mu_r), \label{rgealpha} \\
\frac{d{\overline m}_q(\mu_r)}{d\ln{\mu_r^2}}&=&{\overline m}_q(\mu_r) \gamma_m(a_s)=-{\overline m}_q(\mu_r) \sum_{i=0}^\infty \gamma_i a^{i+1}_s(\mu_r), \label{rgealpha}
\end{eqnarray}
where $a_s(\mu_r)=\alpha_s(\mu_r)/(4\pi)$. The $\{\beta_i\}$-functions and $\{\gamma_i\}$-functions have been calculated up to five-loop level in the $\overline{\rm MS}$-scheme~\cite{Politzer:1973fx, Politzer:1974fr, Gross:1973id, Gross:1973ju, Baikov:2016tgj, Vermaseren:1997fq, Chetyrkin:1997dh, Chetyrkin:2004mf, Baikov:2014qja}, e.g. $\beta_0=11-{2}n_f/{3}$ and $\gamma_0=4$, where $n_f$ is the number of active flavors. Using the reference points, such as $\alpha_s(M_Z)=0.1179\pm0.0009$ and ${\overline m}_t({\overline m}_t)=162.5^{+2.1}_{-1.5}$ GeV given in Particle Data Group~\cite{Workman:2022zbs}, one can obtain their values at any scale.

Because of renormalization group invariance (RGI), the physical observable is independent of any choices of renormalization scheme and renormalization scale. However for a fixed-order pQCD prediction, the mismatching of $\alpha_s$ and the pQCD coefficients at each orders leads to the well-known renormalization scheme-and-scale ambiguities. In order to eliminate such artificially introduced renormalization scheme-and-scale ambiguities, the principle of maximum conformality (PMC)~\cite{Brodsky:2011ta, Brodsky:2011ig, Brodsky:2012rj, Mojaza:2012mf} has been suggested in the literature. The PMC provides the underlying principle for the well-known Brodsky-Lepage-Mackenzie (BLM) method~\cite{Brodsky:1982gc} and provides a rule for generalizing the BLM procedure up to all orders. A short review of the development of PMC from BLM can be found in Ref.\cite{Brodsky:2023iap}. All the features previously observed in the BLM literature are also adaptable to PMC with or without proper transformations, e.g. only the RG-involved $n_f$-terms in the pQCD series should be treated as non-conformal terms and be adopted for setting the correct magnitude of $\alpha_s$. The PMC thus provides a rigorous scale-setting approach for obtaining unambiguous fixed-order pQCD predictions consistent with the principles of renormalization group~\footnote{A practical way of achieving scheme-and-scale invariant prediction directly from the initial series, which is called as principal of minimum sensitivity (PMS)~\cite{Stevenson:1980du, Stevenson:1981vj} has been suggested in the literature. It assumes that all uncalculated higher-order terms give zero contribution and determines the optimal scheme and scale by requiring the slope of the pQCD series over scheme and scale choices vanish. Since the PMS breaks the standard renormalization group invariance~\cite{Wu:2014iba}, it thus cannot be treated as a strict solution of conventional scheme-and-scale ambiguities, which however could be treated as an effective treatment~\cite{Ma:2014oba, Ma:2017xef}.}. And it's prediction satisfies all of the requirements of renormalization group invariance~\cite{Brodsky:2012ms, Wu:2013ei}. The complete discussions about the PMC can be found in review articles~\cite{Wu:2018cmb, Wu:2019mky, DiGiustino:2023jiq}.

Up to now, the PMC method has been successfully applied to many high-energy processes, c.f. Refs.\cite{Du:2018dma, Yu:2018hgw, Yu:2019mce, Huang:2020rtx, Yu:2020tri, Huang:2020skl, Huang:2021kzc}, which aims to determine the correct magnitude of the strong running coupling $\alpha_s$ of the pQCD series by using the procedures suggested in Refs.\cite{Brodsky:2013vpa, Shen:2017pdu}. However, there are also many other progresses associated with both the running coupling $\alpha_s$ and the quark $\overline{\rm MS}$ running mass $\overline{m}_q$. If the pQCD series contains both the $n_f$-terms related to the renormalization of $\alpha_s$ and the $n_f$-terms related to the renormalization of ${\overline m}_q$, some extra treatments must be applied before using the formulas listed in Refs.\cite{Brodsky:2013vpa, Shen:2017pdu}, since the formulas there are based on the assumption that the conformal terms and non-conformal terms have been correctly distributed~\footnote{We have observed that a study on the relation between the OS mass and the $\overline{\rm MS}$ running mass by using the PMC has been issued in Ref.\cite{Salinas-Arizmendi:2022pzj}. It should be pointed out that the authors there do not distinguish the functions of various $n_f$-terms in the perturbative series, which is however important to achieve a precise prediction~\cite{Wang:2013bla}. Thus their treatments can only be treated as an effective but not an strict application of the PMC.}. To achieve the correct PMC prediction, the degeneracy relations among different orders, which are general property of the QCD theory~\cite{Bi:2015wea}, should be applied correctly. In this paper, we will derive new degeneracy relations with the help of the RGEs involving both the $\beta$-function and the quark mass anomalous dimension $\gamma_m$-function, which lead to the improved PMC scale-setting procedures. Such procedures will then be applied for determining the top-quark OS mass by simultaneously fixing the correct magnitudes of the $\alpha_s$ and the quark $\overline{\rm MS}$ running mass ${\overline m}_q$ of the perturbative series with the help of RGEs for either the running coupling $\alpha_s$ or the running mass.

The remaining parts of this paper are organized as follows. In Sec.II, we will show the special degeneracy relations of the non-conformal terms in the perturbative coefficients by using the ${\cal R}_\delta$-scheme. Then we will give the improved procedures of the PMC scale-setting approach under the running mass scheme. We will apply them to determine the top-quark OS mass $M_t$ via its perturbative relation to the $\overline{\rm MS}$ mass in Sec.III. Section IV is reserved for a summary.

\section{Calculation technology \label{II}}

\subsection{Observables in ${\cal R}_\delta$-scheme}

The ${\cal R}_\delta$-scheme represents the ${\rm MS}$-type renormalization scheme with a subtraction term $\ln(4\pi)-\gamma_{E}-\delta$, where the $\delta$ is an arbitrary finite number~\cite{Brodsky:2013vpa}. Especially, ${\cal R}_{\delta=0}=\overline{\rm MS}$. As the extension of Ref.\cite{Brodsky:2013vpa}, we consider the pQCD prediction $\rho(Q^2)$ including the $\overline{\rm MS}$ running mass. In the reference scheme ${\cal R}_0$, it can be written as:
\begin{eqnarray}
\rho_0(Q^2)=r_0 {\overline m}_q^n(\mu_r)\left[1+\sum_{k=1}^\infty r_{k}(\mu_r^2/Q^2)a_s^k(\mu_r)\right], \label{observable} \\ \nonumber
\end{eqnarray}
where $\rho_0$ denotes the pQCD prediction $\rho$ in the ${\cal R}_0$ scheme, $Q$ represents a physical scale of the measured observable~\footnote{The $Q$ represents the typical physical scale of the measured observable, which always appears in the pQCD calculation associated with $\mu_r$. For example, in the calculation of the $n$-loop pQCD correction to the cross section of $e^+(k_1)+e^-(k_2) \to \gamma^* \to q(k_3)+{\bar q}(k_4)$, the calculated results include terms like $\ln(\mu_r^2/s)$ where $s=(k_1+k_2)^2=(k_3+k_4)^2$ represents the squared center-of-mass energy. Thus, the scale $Q$ appears in the expression automatically as $Q=\sqrt{s}$.}, ${\overline m}_q$ is the quark $\overline{\rm MS}$ running mass, $n$ is the power of ${\overline m}_q$ associated with the tree-level term, and the $r_0$ is a global factor. For simplicity, we assume that $r_0$ does not have $a_s$. The pQCD series will be independent of the choice of renormalization scale $\mu_r$, if it has been calculated up to all orders. However, it is not feasible to achieve this goal due to the difficulty of high-order calculation. Generally, the pQCD series becomes renormalization scale and scheme dependent at any finite order, whose dependence can be exposed by using the ${\cal R}_\delta$-scheme. One can derive the general expression for $\rho$ in ${\cal R}_\delta$ by using the scale displacements in any ${\cal R}_\delta$-scheme:
\begin{eqnarray}
a_s(\mu_r) &=& a_s(\mu_\delta) + \sum_{n=1}^\infty \frac{1}{n!} { \frac{{\rm d}^n a_s(\mu_r)}{({\rm d} \ln \mu_r^2)^n}{\bigg |}_{\mu_r=\mu_\delta}(-\delta)^n}, \\
{\overline m}_q(\mu_r) &=& {\overline m}_q(\mu_\delta)+\sum_{n=1}^\infty\frac{1}{n!}\frac{d^n{\overline m}_q(\mu_r)}{(d\ln{\mu_r^2})^n}{\bigg |}_{\mu_r=\mu_\delta}(-\delta)^n, \\ \nonumber
\end{eqnarray}
where $\delta=-\ln{(\mu_r^2/\mu_\delta^2)}$. It is useful to derive the general displacement relations as expansions up to fixed order, which are shown in the Appendix~\ref{A}.

Inserting these scale displacements into Eq.(\ref{observable}), one can obtain the expression of $\rho_\delta$ for an arbitrary $\delta$ in any ${\cal R}_\delta$-scheme. i.e.,
\begin{eqnarray}
\rho_\delta(Q^2)&=&r_0 {\overline m}_q^n(\mu_{\delta})\Big\{1+(r_1+n \gamma_0 \delta)a_s(\mu_{\delta}) \nonumber\\
&+&\Big[r_2+\beta_0r_1\delta+n(\gamma_1+\gamma_0r_1)\delta+\frac{n}{2}\beta_0\gamma_0\delta^2\nonumber\\
&+&\frac{n^2}{2}\gamma_0^2\delta^2\Big]a^2_s(\mu_{\delta})+\mathcal{O}[a^3_s(\mu_{\delta})]\Big\}, \label{rhodeltas}
\end{eqnarray}
where $\rho_\delta$ denotes the pQCD prediction $\rho$ in the ${\cal R}_\delta$ scheme and $\mu_{\delta}^2=\mu_r^2 e^\delta$. The useful expression of $\rho_\delta$ up to $a_s^4$-order is given in the Appendix~\ref{B}. It is easy to go back to expression $\rho_0$ by taking $\delta=0$. More description about the ${\cal R}_\delta$-scheme can be found in Sec. II of Ref.\cite{Brodsky:2013vpa}.

The renormalization group invariance requires that the perturbative series up to all orders for a physical observable is independent of theoretical convention; i.e.,
\begin{eqnarray}
\frac{d \rho_{\delta}}{d \delta}&=&\frac{\partial \rho_{\delta}}{\partial \delta}+\beta(a_s)\frac{\partial \rho_{\delta}}{\partial a_s}+{\overline m}_q \gamma_m(a_s)\frac{\partial \rho_{\delta}}{\partial {\overline m}_q}. \nonumber\\
&=&0.
\end{eqnarray}
Thus, we can obtain
\begin{eqnarray}
\frac{\partial \rho_{\delta}}{\partial \delta}&=&-\beta(a_s)\frac{\partial \rho_{\delta}}{\partial a_s}-{\overline m}_q \gamma_m(a_s)\frac{\partial \rho_{\delta}}{\partial {\overline m}_q}.
\end{eqnarray}
Therefore, by absorbing all $\{\beta_i\}$ and $\{\gamma_i\}$ dependence into the running coupling constant and quark running mass in Eq.(\ref{rhodeltas}), one can obtain a scheme-invariant prediction due to the $\delta$-dependence is vanished. The coefficients of the resultant series will thus be equal to those of the conformal (or scale-invariant) theory, e.g. $\partial \rho_\delta /\partial \delta = 0$.

The expression in Eq.(\ref{rhodeltas}) also exposes the pattern of $\{\beta_i\}$-terms and $\{\gamma_i\}$-terms in the coefficients at each order. Since there is nothing special about any particular value of $\delta$, it is possible to infer some degenerate relations between certain coefficients of the $\{\beta_i, \gamma_i\}$-terms from the expression $\rho_\delta$. That is, the coefficients of $\beta_0 a_s^2$ and $n\gamma_0 a_s^2$ can be set equal, since their coefficients are both $r_1\delta$. Therefore, for any scheme, the expression for $\rho$ can be transformed into a form similar to that of $\rho_\delta$. i.e.,
\begin{eqnarray}
\rho(Q^2)&=&r_0 {\overline m}_q^n(\mu_r)\Big\{1+\Big({\hat r}_{1,0}+n\gamma_0\ln\frac{\mu_r^2}{Q^2}\Big) a_s(\mu_r) \nonumber \\
&+&\Big[{\hat r}_{2,0}+\beta_0{\hat r}_{2,1}+n\gamma_0{\hat r}_{2,1}\nonumber \\
&+&\Big(\beta_0{\hat r}_{1,0}+n\gamma_1+n\gamma_0{\hat r}_{1,0}\Big)\ln\frac{\mu_r^2}{Q^2}\nonumber \\
&+&\frac{1}{2}\Big(n\beta_0\gamma_0+n^2\gamma_0^2\Big)\ln^2\frac{\mu_r^2}{Q^2}\Big] a^2_s(\mu_r)\nonumber \\
&+&\mathcal{O}[a^3_s(\mu_r)]\Big\}, \label{ruij}
\end{eqnarray}
where ${\hat r}_{i,j}$ are coefficients independent of $\mu_r$, ${\hat r}_{i,0}$ are conformal coefficients, and the $\{\beta_i, \gamma_i\}$-terms are non-conformal terms. The useful expression up to $a_s^4$-order is given in the Appendix~\ref{C}. It is easy to find the relationships between the coefficients $r_{k}(\mu_r^2/Q^2)$ and the coefficients ${\hat r}_{i,j}$, i.e.,
\begin{eqnarray}
r_{1}(\mu_r^2/Q^2)&=& {\hat r}_{1,0}+n\gamma_0\ln\frac{\mu_r^2}{Q^2}, \\
r_{2}(\mu_r^2/Q^2)&=& {\hat r}_{2,0}+\beta_0{\hat r}_{2,1}+n\gamma_0{\hat r}_{2,1}\nonumber \\
&+&\Big(\beta_0{\hat r}_{1,0}+n\gamma_1+n\gamma_0{\hat r}_{1,0}\Big)\ln\frac{\mu_r^2}{Q^2}\nonumber \\
&+&\frac{1}{2}\Big(n\beta_0\gamma_0+n^2\gamma_0^2\Big)\ln^2\frac{\mu_r^2}{Q^2},
\end{eqnarray}
The relationships between the coefficients $r_{k}(\mu_r^2/Q^2)(k=3,4)$ and the coefficients ${\hat r}_{i,j}$ are given in the Appendix~\ref{D}. These relationships lead to the systematic procedures to determine the coefficients ${\hat r}_{i,j}$. In some cases, the coefficients $r_{k}(\mu_r^2/Q^2)$ of Eq.(\ref{observable}) are computed numerically, and the $\{\beta_i, \gamma_i\}$ dependence is not known explicitly. However, it is straightforward to obtain the dependence on the number of quark flavors $n_f$, since $n_f$ enters analytically in any loop diagram computation. To applying the PMC scale-setting approach, one should put the pQCD expression into the form of Eq.(\ref{ruij}). Due to the special degeneracy relations in the coefficients of$\{\beta_i, \gamma_i\}$-terms, the $n_f$ series can be matched to the ${\hat r}_{i,j}$ coefficients in a unique way. The $k_{\rm th}$-order coefficient in pQCD has an expansion in $n_f$, which reads
\begin{eqnarray}
r_{k}(\mu_r^2/Q^2)=c_{k,0}+c_{k,1}n_f+...+c_{k,k-1}n_f^{k-1},
\end{eqnarray}
where the coefficients $c_{k,l}$ are obtained from the pQCD calculation and they are function of $\mu_r$ and $Q$. Then, the coefficients ${\hat r}_{i,j}$ of Eq.(\ref{ruij}) can be determined by using its relationship with $r_{k}(\mu_r^2/Q^2)$ and the known coefficients $c_{k,l}$. The detail steps are shown in the Appendix~\ref{E}. In the next section, we will show the improved PMC formulas under the running mass scheme.

\subsection{PMC single-scale approach under the running mass scheme }

Adopting to the PMC single-scale approach~\cite{Shen:2017pdu}, the overall effective running coupling $a_s(Q_*)$ and the effective running mass ${\overline m}_q(Q_*)$ can be formed by absorbing all the non-conformal terms. Eq.(\ref{ruij}) will change to the following conformal series,
\begin{eqnarray}
\rho(Q^2)&=&r_0 {\overline m}_q^n(Q_*)\Big\{1+{\hat r}_{1,0}a_s(Q_*) + {\hat r}_{2,0}a_s^2(Q_*)\nonumber \\
&+&{\hat r}_{3,0}a_s^3(Q_*)+{\hat r}_{4,0}a_s^4(Q_*)+\mathcal{O}[a^5_s(Q_*)]\Big\},
\end{eqnarray}
where $Q_*$ is the PMC scale. More explicitly, by using the scale displacement relations to shift the scale $\mu_r$ to $Q_*$ in Eq.(\ref{ruij}), the PMC scale $Q_*$ can be determined by requiring all non-conformal terms (NonConf.) vanish, i.e.,
\begin{eqnarray}
\rho(Q^2)_{\rm NonConf.}&=&r_0 {\overline m}_q^n(Q_*)\Big\{r_{\rm 1, NonConf.}(Q_*) a_s(Q_*) \nonumber \\
&+&r_{\rm 2, NonConf.}(Q_*) a^2_s(Q_*)\nonumber \\
&+&r_{\rm 3, NonConf.}(Q_*) a^3_s(Q_*)\nonumber \\
&+&r_{\rm 4, NonConf.}(Q_*) a^4_s(Q_*)+{\mathcal O}[a^5_s(Q_*)]\Big\} \nonumber \\
&=&0, \label{nonconf}
\end{eqnarray}
where
\begin{eqnarray}
r_{\rm 1, NonConf.}(Q_*)&=&n\gamma_0\ln\frac{Q_*^2}{Q^2}, \\
r_{\rm 2, NonConf.}(Q_*)&=&\beta_0{\hat r}_{2,1}+n\gamma_0{\hat r}_{2,1}\nonumber \\
&+&\Big(\beta_0{\hat r}_{1,0}+n\gamma_1+n\gamma_0{\hat r}_{1,0}\Big)\ln\frac{Q_*^2}{Q^2}\nonumber \\
&+&\frac{1}{2}\Big(n\beta_0\gamma_0+n^2\gamma_0^2\Big)\ln^2\frac{Q_*^2}{Q^2},
\end{eqnarray}
and the higher-order coefficients $r_{\rm i, NonConf.}(i=3,4)$ are given in the Appendix~\ref{F}.

Due to its perturbative nature, the solution of $\ln(Q^2_*/Q^2)$ can be expand as a power series over $a_s(Q_*)$; i.e.,
\begin{eqnarray}
\ln\frac{Q^2_*}{Q^2}=\sum_{i=0}^{n} S_{i}a^i_s(Q_*),  \label{qstar1}
\end{eqnarray}
where $S_i$ are perturbative coefficients that can be derived by solving the Eq.(\ref{nonconf}). For $n\neq0$, the first three coefficients $S_i(i=0,1,2)$ are
\begin{eqnarray}
S_0&=&0,\\
S_1&=&-\frac{(\beta_0+n\gamma_0){\hat r}_{2,1}}{n\gamma_0}, \\
S_2&=&{\hat r}_{1,0}{\hat r}_{2,1}-{\hat r}_{3,1}-\frac{n\gamma_0{\hat r}_{3,2}}{2}-\frac{\beta_1{\hat r}_{2,1}}{n\gamma_0}\nonumber\\
&+&\beta_0\left(\frac{\gamma_1{\hat r}_{2,1}}{n\gamma_0^2}+\frac{2{\hat r}_{1,0}{\hat r}_{2,1}-2{\hat r}_{3,1}}{n\gamma_0}-\frac{3{\hat r}_{3,2}}{2}\right)\nonumber\\
&+&\beta_0^2\left(\frac{{\hat r}_{1,0}{\hat r}_{2,1}}{n^2\gamma_0^2}-\frac{{\hat r}_{3,2}}{n\gamma_0}\right),
\end{eqnarray}
and the coefficients $S_3$ is shown in the Appendix~\ref{G}. Following the idea, the PMC scale $Q_*$ can be fixed at any order; the correct magnitudes of the quark running mass ${\overline m}_q$ and running coupling constant $a_s$ are determined simultaneously. Matching with the $\mu_r$-independent conformal coefficients ${\hat r}_{i,0}$, the resultant PMC series will be free of conventional renormalization scale ambiguity. In the following section, we will apply these formulas to determine the top-quark OS mass $M_t$ via its perturbative relation to the $\overline{\rm MS}$ mass.

\section{Numerical results \label{III}}

To do the numerical calculation, we adopt $\alpha_s(M_Z)=0.1179\pm0.0009$ and ${\overline m}_t({\overline m}_t)=162.5^{+2.1}_{-1.5}$ GeV~\cite{Workman:2022zbs}. And the running of the strong coupling $\alpha_s(\mu_r)$ is evaluated by using the RunDec program~\cite{Herren:2017osy}.

\subsection{Properties of the top-quark on-shell mass $M_t$}

The relation between the $\overline{\rm MS}$ quark mass and OS quark mass can be written as
\begin{eqnarray}
&&\frac{{\overline m}_t(\mu_r)}{M_t}=\frac{Z_m^{\rm OS}}{Z_m^{\overline{\rm MS}}}=\sum_{n\geq0}a^n_s(\mu_r) z^{(n)}_m(\mu_r),\label{zmmur}
\end{eqnarray}
where the $z^{(0)}_m(\mu_r)=1$ and $z^{(n)}_m(\mu_r)$ is a function of $\ln(\mu_r^2/M_t^2)$. As an expansion of the previous work~\cite{Huang:2020rtx}, we focus on the inverted relation to Eq.(\ref{zmmur}),
\begin{eqnarray}
&&\frac{M_t}{{\overline m}_t(\mu_r)}=\frac{Z_m^{\overline{\rm MS}}}{Z_m^{\rm OS}}=\sum_{n\geq0}a^n_s(\mu_r) c^{(n)}_m(\mu_r),
\end{eqnarray}
where the $c^{(0)}_m(\mu_r)=1$ and $c^{(n)}_m(\mu_r)$ is a function of $\ln(\mu_r^2/{\overline m}_t^2(\mu_r))$. Then, we can determine the top-quark OS mass using the following relationship, i.e.,
\begin{eqnarray}
M_t&=&{\overline m}_t(\mu_r)\Big\{1+r_1(\mu_r)a_s(\mu_r) + r_2(\mu_r)a_s^2(\mu_r)\nonumber \\
&+&r_3(\mu_r)a_s^3(\mu_r)+ r_4(\mu_r)a_s^4(\mu_r)+\mathcal{O}[a^5_s(\mu_r)]\Big\}, \label{relation}
\end{eqnarray}
where the perturbative coefficients $r_i(\mu_r)(i=1,2,3,4)$ have been known in Refs.\cite{Marquard:2015qpa, Marquard:2016dcn} and $r_i(\mu_r)$ are functions of $\ln (\mu_r^2/{\overline m}_t^2(\mu_r))$. The $\mu_r$ dependence of the top-quark $\overline{\rm MS}$ running mass ${\overline m}_t(\mu_r)$ is governed by the quark mass anomalous dimension $\gamma_m$, which has been calculated up to ${\mathcal O}(\alpha_s^5)$~\cite{Chetyrkin:2004mf, Baikov:2014qja}. Then the top-quark $\overline{\rm MS}$ running mass ${\overline m}_t(\mu_r)$ can be fixed by the following equation~\cite{Baikov:2014qja}:
\begin{eqnarray}
{\overline m}_t(\mu_r)={\overline m}_t({\overline m}_t)\frac{c_t[\alpha_s(\mu_r)/\pi]}{c_t[\alpha_s({\overline m}_t)/\pi]}, \label{mtrun}
\end{eqnarray}
where the function $c_t[x]=x^{4/7}(1+1.19796x+1.79348x^2-0.683433x^3-3.53562x^4)$.

\begin{figure}[htb]
\centering
\includegraphics[width=0.45\textwidth]{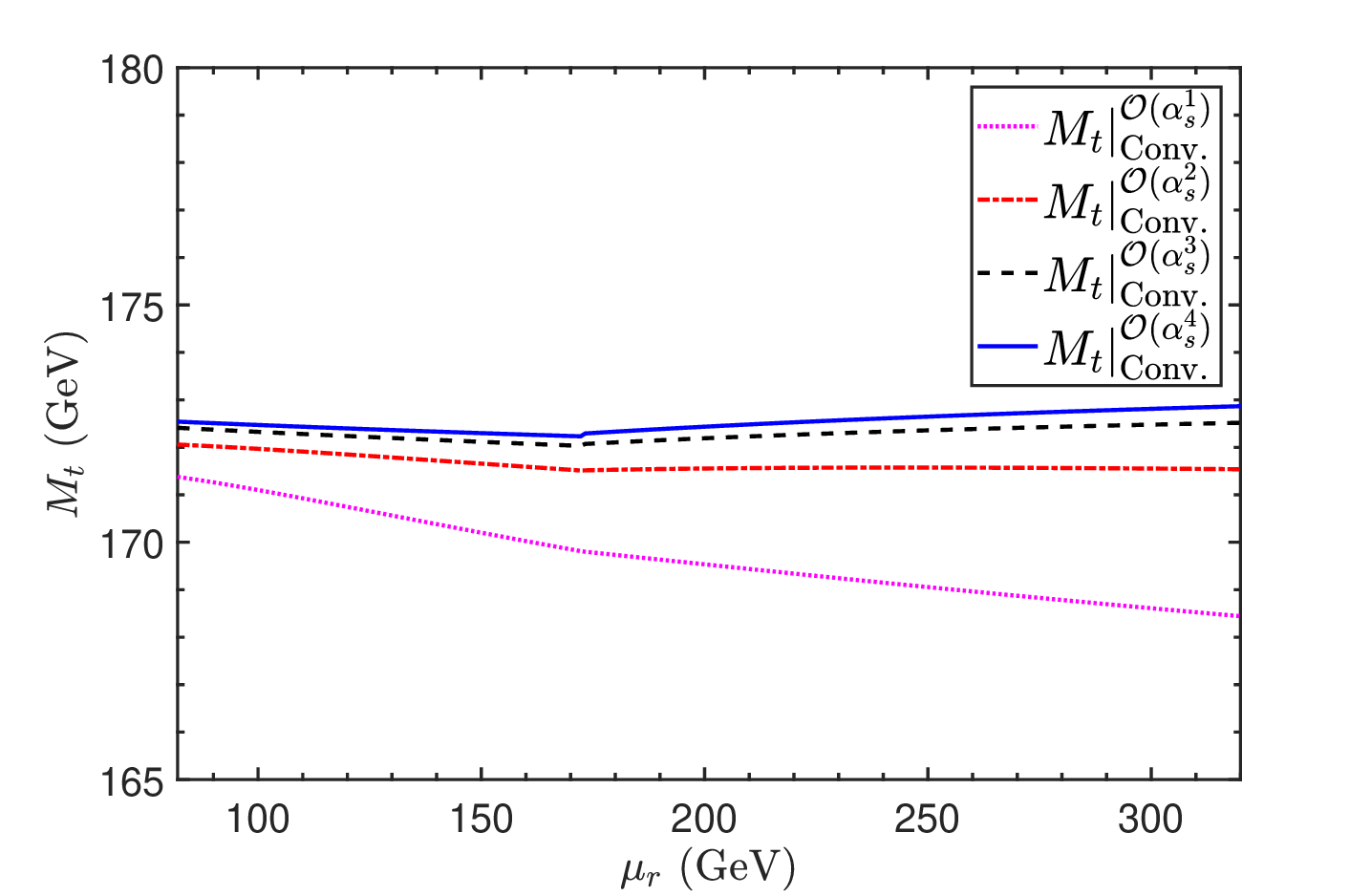}
\caption{The top-quark OS mass $M_t$ versus the renormalization scale ($\mu_r$) under conventional scale-setting approach up to different perturbative orders.} \label{Convurdepen}
\end{figure}

By taking all input parameters to be their central values, we show the top-quark OS mass $M_t$ under conventional scale-setting approach in FIG.~\ref{Convurdepen}. FIG.~\ref{Convurdepen} shows that in agreement with the conventional wisdom, the renormalization scale dependence of conventional series becomes smaller when more loop terms have been included. Numerically, we have $M_t|_{\rm Conv.}^{\mathcal{O}(\alpha_s^4)} =[172.23, 172.88]$ for $\mu_r \in [{\overline m}_t({\overline m}_t)/2, 2{\overline m}_t({\overline m}_t)]$; e.g., the net scale uncertainty becomes $\sim0.4\%$ for a $\alpha_s^4$-order correction. Such small net scale dependence for the prediction up to $a_s^4$-order is due to the well convergent behavior of the perturbative series. The relative magnitudes among different orders change greatly for different choice of $\mu_r$, for example, the relative magnitudes of the leading-order-terms (LO): next-to-leading-order-terms (NLO): next-to-next-to-leading-order-terms (N$^2$LO): next-to-next-to-next-to-leading-order-terms (N$^3$LO): next-to-next-to-next-to-next-to-leading-order-terms (N$^4$LO) $\simeq$ 1: $4.60\%$: $0.98\%$: $0.30\%$: $0.12\%$ for $\mu_r={\overline m}_t({\overline m}_t)$, which represents good perturbative nature. More specifically, the $M_t$ up to N$^4$LO-level has the following perturbative behavior:
\begin{eqnarray}
M_t|_{\rm Conv.}&=&162.5^{-8.15}_{-0.83} +7.48^{+6.36}_{+0.62} + 1.60^{+1.80}_{+0.14}\nonumber \\
&&+ 0.49^{+0.47}_{+0.03} + 0.19^{+0.14}_{+0.01}\nonumber \\
&=&172.26^{+0.62}_{-0.03} ~(\rm GeV), \label{conventionscale}
\end{eqnarray}
whose central values are for $\mu_r={\overline m}_t({\overline m}_t)$, and the scale uncertainties are estimated by varying $\mu_r \in [{\overline m}_t({\overline m}_t)/2, 2{\overline m}_t({\overline m}_t)]$. The higher-order prediction of $M_t$ is not a monotonic function of $\mu_r$, which leads to the asymmetric uncertainty. By using another usual choice of $\mu_r=172.5$ GeV and vary it within the range of $[1/2 \times 172.5, 2 \times 172.5]$, we obtain $M_t=172.29^{+0.64}_{-0.06}$ GeV. The central value shifts from $172.26$ GeV by $+0.03$ GeV, and its uncertainty remains asymmetric.

\begin{figure}[htb]
\includegraphics[width=0.48\textwidth]{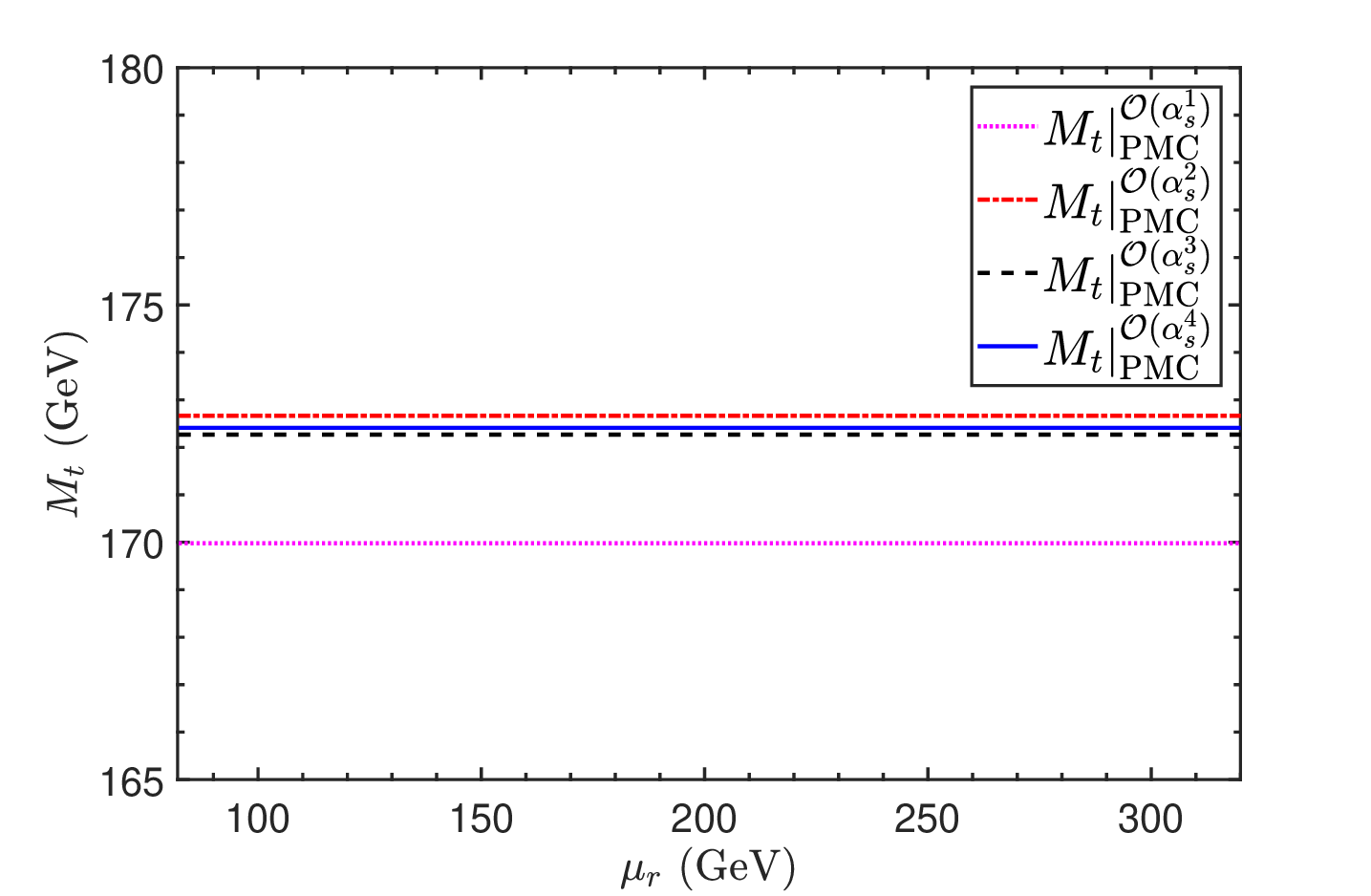}
\caption{The top-quark OS mass $M_t$ versus the renormalization scale ($\mu_r$) under PMC single-scale approach up to different perturbative orders.} \label{pmcurdepen}
\end{figure}

We present $M_t$ under PMC scale-setting approach in FIG.~\ref{pmcurdepen}, which shows the top-quark OS mass $M_t$ under PMC single-scale approach up to different perturbative orders, i.e.,
\begin{eqnarray}
M_t|_{\rm PMC}^{\mathcal{O}(\alpha_s^n)}=\{170.01, 172.66, 172.27, 172.41\} ~(\rm GeV)
\end{eqnarray}
for $n=1$, $2$, $3$ and $4$, respectively. There is no renormalization scale ambiguity for the PMC prediction, and $M_t$ shows a quicker trends of approaching its steady value. After applying the PMC, the perturbative nature of the $M_t$ pQCD series is greatly improved due to the elimination of divergent renormalon terms, and the relative importance of the LO-terms: NLO-terms: N$^2$LO-terms: N$^3$LO-terms: N$^4$LO-terms changes to $1$: $4.79\%$: $-1.15\%$: $-0.15\%$: $0.07\%$. Up to N$^4$LO-level, we have
\begin{eqnarray}
M_t|_{\rm PMC}&=&{\overline m}_t(Q_*)\Big\{1+{\hat r}_{1,0}a_s(Q_*) + {\hat r}_{2,0}a_s^2(Q_*)\nonumber \\
&&+{\hat r}_{3,0}a_s^3(Q_*)+ {\hat r}_{4,0}a_s^4(Q_*)+\mathcal{O}[a^5_s(Q_*)]\Big\}\nonumber \\
&=&166.49 + 7.97 - 1.92 - 0.25 + 0.12\nonumber \\
&=&172.41 ~(\rm GeV), \label{PMCseries}
\end{eqnarray}
where the PMC scale $Q_*$ can be fixed up to next-to-next-to-leading-log (N$^2$LL) accuracy by using Eq.(\ref{qstar1}); i.e.,
\begin{eqnarray}
\ln\frac{Q^2_*}{{\overline m}_t^2(Q_*)}&=&-68.73 a_s(Q_*)+247.483 a^2_s(Q_*)\nonumber \\
&&-6447.27 a^3_s(Q_*), \label{PMCscaleQstar}
\end{eqnarray}
which leads to $Q_*=123.3$ GeV. Due to its perturbative nature, we take the magnitude of the last known term as the unknown N$^3$LL term as a conservative estimation of the unknown perturbative terms. We then obtain a scale shift of $\Delta Q_*=\left(^{+0.2}_{-0.3}\right)$ GeV, which leads to
\begin{eqnarray}
\Delta M_t|_{\rm PMC}=(^{+0.04}_{-0.03}) ~(\rm GeV).  \label{residualscale}
\end{eqnarray}
This uncertainty is called as {\it the first kind of residual scale dependence due to unknown higher-order terms}~\cite{Zheng:2013uja}. Such residual scale dependence is distinct from the conventional scale ambiguities and is suppressed due to the perturbative nature of the PMC scale. For the present case, the residual scale dependence (\ref{residualscale}) is much smaller than conventional scale uncertainty (\ref{conventionscale}).

\subsection{Theoretical uncertainties}

In the pQCD calculation, the magnitude of unknown perturbative terms is also a major source of uncertainties. It is helpful to find a reliable prediction of the unknown higher-order terms. The Pad\'{e} approximation approach (PAA)~\cite{Basdevant:1972fe, Samuel:1992qg} is a well known method to estimate the $(n+1)_{\rm th}$-order coefficient for a given $n_{\rm th}$-order perturbative series, which has been tested on various known QCD results~\cite{Samuel:1995jc}.
More explicitly, for the pQCD approximant $\rho(Q^2)=c_1 a_s+c_2 a_s^2+c_3 a_s^3+c_4 a_s^4$, the preferable [$n$/$n$+1]-type PAA predictions~\cite{Gardi:1996iq} of the $a_s^5$-terms of $M_t$ under the conventional scale-setting approach is
\begin{eqnarray}
\rho^{\rm N^5LO}_{[1/2]}&=&\frac{2c_2 c_3 c_4-c_3^3-c_1 c_4^2} {c_2^2-c_1 c_3} a_s^5,
\end{eqnarray}
and the preferable [0/$n$-1]-type PAA predictions the $a_s^5$-terms of $M_t$ under the PMC scale-setting approach~\cite{Du:2018dma} is
\begin{eqnarray}
\rho^{\rm N^5LO}_{[0/3]}&=&\frac{c_2^4-3c_1 c_2^2 c_3+2c_1^2 c_2 c_4+c_1^2 c_3^2} {c_1^3} a_s^5.
\end{eqnarray}
This uncertainty could be called as {\it the second kind of residual scale dependence} due to unknown higher-order terms. Then, our prediction of the magnitude of the N$^5$LO-terms of top-quark OS mass $M_t$ is
\begin{eqnarray}
\Delta M_t|_{\rm Conv.}^{\rm N^5LO}&=&\pm0.08 ~(\rm GeV), \\
\Delta M_t|_{\rm PMC}^{\rm N^5LO}&=&\pm0.02 ~(\rm GeV).
\end{eqnarray}
Note that, the conventional result is obtained by taking $\mu_r={\overline m}_t({\overline m}_t)$, whose numerical result will vary with the renormalization scale due to the coefficients $c_i$ are not fixed at each order. However, the PAA prediction of the PMC series has no renormalization scale ambiguity, since the PMC coefficients are scale-invariant.

The combination of the above two residual scale dependence leads to a net perturbative uncertainty due to uncalculated higher-order terms under conventional and PMC scale-setting approaches; i.e.,
\begin{eqnarray}
\Delta M_t|_{\rm Conv.}^{\rm High~order}&=&(^{+0.63}_{-0.09}) ~(\rm GeV), \\
\Delta M_t|_{\rm PMC}^{\rm High~order}&=&\pm0.04 ~(\rm GeV).
\end{eqnarray}
This shows that the PMC scale-invariant series provides a more accurate basis than the conventional scale-dependent one for estimating the uncertainty caused by uncalculated higher-order terms.

There are uncertainties from $\Delta\alpha_s(M_Z)$ and $\Delta{\overline m}_t({\overline m}_t)$. As an estimation, by using $\Delta\alpha_s(M_Z)=\pm0.0009$~\cite{Workman:2022zbs}, we obtain
\begin{eqnarray}
\Delta M_t|_{\rm Conv.}^{\Delta\alpha_s(M_Z)}&=&\pm0.09 ~(\rm GeV), \\
\Delta M_t|_{\rm PMC}^{\Delta\alpha_s(M_Z)}&=&(^{+0.10}_{-0.08}) ~(\rm GeV).
\end{eqnarray}
And for estimating the uncertainty from the top-quark $\overline{\rm MS}$ mass, by taking $\Delta{\overline m}_t({\overline m}_t)=(^{+2.1}_{-1.5})$ GeV, we obtain
\begin{eqnarray}
\Delta M_t|_{\rm Conv.}^{\Delta{\overline m}_t({\overline m}_t)}&=&(^{+2.20}_{-1.57}) ~(\rm GeV), \\
\Delta M_t|_{\rm PMC}^{\Delta{\overline m}_t({\overline m}_t)}&=&(^{+2.21}_{-1.57}) ~(\rm GeV).
\end{eqnarray}
This shows that the magnitude of $\Delta M_t$ is close to that of $\Delta{\overline m}_t({\overline m}_t)$.

We then obtain the final result of top-quark OS mass,
\begin{eqnarray}
M_t|_{\rm Conv.}&=&172.26^{+2.29}_{-1.58} ~(\rm GeV), \\
M_t|_{\rm PMC}&=&172.41^{+2.21}_{-1.57} ~(\rm GeV),
\end{eqnarray}
where the uncertainties are squared average of those from $\Delta M_t|^{\text{High order}}$, $\Delta M_t|^{\Delta\alpha_s(M_Z)}$, and $\Delta M_t|^{\Delta{\overline m}_t({\overline m}_t)}$, respectively. Among the uncertainties, the one caused by $\Delta{\overline m}_t({\overline m}_t)$ is dominant~\footnote{If the value of ${\overline m}_t({\overline m}_t)$ can be measured very accurately to avoid the $\Delta {\overline m}_t({\overline m}_t)$ uncertainty, we will obtain $M_t|_{\rm Conv.}=172.26^{+0.64}_{-0.13} (\rm GeV)$, and $M_t|_{\rm PMC}=172.41^{+0.11}_{-0.09} (\rm GeV)$.}, and we need more accurate data to suppress this uncertainty.

\begin{table}[htb]
\centering
\begin{tabular}{c c}
\hline
  & $M_t$~(\rm GeV) \\
\hline
 PMC (this work) & $172.41^{+2.21}_{-1.57}$ \\
\hline
 Conv. (this work) & $172.26^{+2.29}_{-1.58}$ \\
\hline
 CMS~\cite{CMS:2022emx} & $172.93\pm1.36$ \\
\hline
 PDG~\cite{Workman:2022zbs} & $172.5\pm0.7$ \\
\hline
 ATLAS and CMS~\cite{ATLAS:2022aof} & $173.4^{+1.8}_{-2.0}$ \\
\hline
  \cite{Wang:2020mel} & $172.5\pm1.4$ \\
\hline
  \cite{Garzelli:2023rvx} & $171.54^{+0.28}_{-0.31}$ \\
\hline
 \cite{Cridge:2023ztj} & $173.0\pm0.6$ \\
\hline
\end{tabular}
\caption{A comparison of the top-quark OS mass $M_t$ under various approaches. Refs.\cite{Wang:2020mel, Garzelli:2023rvx, Cridge:2023ztj} derive the magnitude of the OS mass via a comparison of experimental data with the theoretical predictions on the top-pair production cross-sections at the hadronic colliders. }  \label{osmassC}
\end{table}

We compare our results with the experimental measurements~\cite{CMS:2022emx, Workman:2022zbs, ATLAS:2022aof} and some other theoretical predictions based on the analyses of top-pair production at the hadronic colliders~\cite{Garzelli:2023rvx, Cridge:2023ztj,Wang:2020mel} in Table~\ref{osmassC}. All the predictions are consistent with each other within reasonable errors. And due to large input error of the top-quark $\overline{\rm MS}$ mass ${\overline m}_t({\overline m}_t)$, our results show larger uncertainty than the results of Refs.\cite{Garzelli:2023rvx, Cridge:2023ztj,Wang:2020mel}. Up to the present known N$^4$LO level, the predictions under the PMC and conventional scale-setting approaches are both consistent with the newest experimental measurement~\cite{CMS:2022emx}.

At present, the experimentally measured value of top-quark OS mass is more precise than that of the top-quark $\overline{\rm MS}$ mass, one can extract the $\overline{\rm MS}$ mass ${\overline m}_t({\overline m}_t)$ by inversely using Eq.(\ref{relation}) or the resultant PMC series. We will show such discussion in the following section.

\subsection{Extracting the top-quark $\overline{\rm MS}$ mass ${\overline m}_t({\overline m}_t)$}

It is also a useful way to extract the $\overline{\rm MS}$ mass ${\overline m}_t({\overline m}_t)$ from the experimental result of OS mass. The PDG~\cite{Workman:2022zbs} shows the world average result of top-quark OS mass is $M_t=172.5\pm0.7$ GeV, whose accuracy is more precise than that of the top-quark $\overline{\rm MS}$ mass, e.g. ${\overline m}_t({\overline m}_t)=162.5^{+2.1}_{-1.5}$ GeV.

In the conventional scale-setting approach, one can extract the $\overline{\rm MS}$ mass ${\overline m}_t({\overline m}_t)$ by using the Eq.(\ref{relation}) and setting the renormalization scale $\mu_r={\overline m}_t({\overline m}_t)$; i.e.,
\begin{eqnarray}
{\overline m}_t({\overline m}_t)|_{\rm Conv.}=162.73^{+0.67}_{-0.67} ~(\rm GeV),
\end{eqnarray}
where the central value is obtained by adopting $M_t=172.5$ GeV and the uncertainty is caused by the $\Delta M_t=\pm0.7$ GeV.

If considering the uncertainty of the renormalization scale $\mu_r \in[\frac{1}{2}{\overline m}_t({\overline m}_t),2{\overline m}_t({\overline m}_t)]$, we obtain
\begin{eqnarray}
{\overline m}_t({\overline m}_t)|_{\rm Conv.}=162.73^{+0.00}_{-1.16} ~(\rm GeV),
\end{eqnarray}
Thus, the conventional prediction is
\begin{eqnarray}
{\overline m}_t({\overline m}_t)|_{\rm Conv.}=162.73^{+0.67}_{-1.34} ~(\rm GeV), \label{convmt}
\end{eqnarray}
where the uncertainty is squared average of those from $\Delta M_t=\pm0.7$ GeV and the renormalization scale $\mu_r \in[\frac{1}{2}{\overline m}_t({\overline m}_t),2{\overline m}_t({\overline m}_t)]$.

In the PMC scale-setting approach, one can extract the $\overline{\rm MS}$ mass ${\overline m}_t({\overline m}_t)$ by using the PMC series without renormalization scale uncertainty; i.e.,
\begin{eqnarray}
M_t&=&{\overline m}_t(Q_*)\Big\{1+{\hat r}_{1,0}a_s(Q_*) + {\hat r}_{2,0}a_s^2(Q_*) \nonumber \\
&&+{\hat r}_{3,0}a_s^3(Q_*)+ {\hat r}_{4,0}a_s^4(Q_*)+\mathcal{O}[a^5_s(Q_*)]\Big\},
\end{eqnarray}
where the PMC scale $Q_*$ satisfies the Eq.(\ref{PMCscaleQstar}). Thus, the PMC prediction is
\begin{eqnarray}
{\overline m}_t({\overline m}_t)|_{\rm PMC}=163.08^{+0.66}_{-0.66} ~(\rm GeV), \label{PMCmt}
\end{eqnarray}
where the central value is obtained by setting $M_t=172.5$ GeV and the uncertainty is caused by $\Delta M_t=\pm0.7$ GeV. It can be found that the PMC result predicted by Eq.(\ref{PMCmt}) is more accurate than the conventional result predicted by Eq.(\ref{convmt}). Finally, the PMC result and the conventional result both overlap with our previous work ${\overline m}_t({\overline m}_t)=162.6^{+0.4}_{-0.4}$ GeV that has been done by using the RGE of $\alpha_s$ alone~\cite{Huang:2020rtx} and the experimental result ${\overline m}_t({\overline m}_t)=162.9\pm0.5\pm1.0^{+2.1}_{-1.2}$ GeV~\cite{ATLAS:2019guf}.

\section{Summary}

In the present paper, we have derived new degeneracy relations with the help of the RGEs involving both the $\beta$-function and the quark mass anomalous dimension $\gamma_m$-function, which lead to the improved PMC scale-setting procedures. Such procedures can be used to eliminate the conventional renormalization scale ambiguity of the fixed-order pQCD series under the $\overline{\rm MS}$ running mass scheme, which simultaneously determines the correct magnitudes of the $\alpha_s$ and the $\overline{\rm MS}$ running mass ${\overline m}_q$ of the perturbative series with the help of RGEs. By applying such formulas, we have obtained a scale invariant $\overline{\rm MS}$-on-shell quark mass relation. Consequently, we determine the top-quark on-shell or $\overline{\rm MS}$ mass without conventional renormalization scale ambiguity. By taking the top-quark $\overline{\rm MS}$ mass ${\overline m}_t({\overline m}_t)=162.5^{+2.1}_{-1.5}$ given in PDG as an input, we obtain the top-quark OS mass $M_t|_{\rm PMC}=172.41^{+2.21}_{-1.57} ~(\rm GeV)$. It can be found that such result has a larger uncertainty than the experimental value, since the input error $\Delta{\overline m}_t({\overline m}_t)$ still has a somewhat larger magnitude. We have also inversely determined the top-quark $\overline{\rm MS}$ mass ${\overline m}_t({\overline m}_t)=163.08^{+0.66}_{-0.66}$ by using the top-quark OS mass $M_t=172.5\pm0.7$ GeV as input; The resultant prediction of the top-quark $\overline{\rm MS}$ mass is more precise than the experimental value given in PDG.

The accuracy of the pQCD prediction under the $\overline{\rm MS}$ running mass scheme depends strongly on the exact values of $\alpha_s$ and ${\overline m}_q$. After applying the PMC, the correct values of the effective $\alpha_s$ and ${\overline m}_q$ can be determined simultaneously, its resultant more convergent pQCD series also leads to a much smaller residual scale dependence, thus a reliable and precise pQCD prediction can be achieved. It is also a very useful method to determine the bottom-quark OS mass and the charm-quark OS mass in the future.

\hspace{2cm}

\acknowledgments
We thank Sheng-Quan Wang, Jian-Ming Shen, Jun Zeng and Qing Yu for helpful discussions. This work was supported in part by the National Natural Science Foundation of China under Grant No.12247129, No.12175025, and No.12347101, and by the Graduate Research and Innovation Foundation of Chongqing, China under Grant No.ydstd1912.


\begin{widetext}

\appendix

\section{Scale Displacement Relation } \label{A}

The general scale displacement relations of the strong coupling constant $a_s$ and the quark running mass $\overline{m}_q$ up to fourth-order are
\begin{eqnarray}
a^k_s(\mu_r) &=& a^k_s(\mu_\delta) + k \beta_0 \delta a^{k+1}_s(\mu_\delta) + k \bigg(\beta_1 \delta + \frac{k+1}{2}\beta_0^2 \delta^2 \bigg) a^{k+2}_s(\mu_\delta) \delta+ k\bigg[\beta_2 \nonumber\\
&+&\frac{2k+3}{2}\beta_0\beta_1 \delta^2 +\frac{(k+1)(k+2)}{3!}\beta_0^3 \delta^3 \bigg]a^{k+3}_s(\mu_\delta)+\mathcal{O}[a^{k+4}_s(\mu_\delta)]. \label{scaledis} \\
{\overline m}_q^n(\mu_r)&=&{\overline m}_q^n(\mu_\delta)\bigg\{1+n\gamma_0\delta a_s+\bigg[\frac{1}{2} \left(n\beta_0\gamma_0+n^2\gamma_0^2\right)\delta^2
+n\gamma_1\delta\bigg]a_s^2 +\bigg[n\gamma_2\delta+\frac{1}{2} (2n\beta_0\gamma_1\nonumber \\
&+&n\beta_1\gamma_0+2n^2\gamma_0\gamma_1)\delta^2+\frac{1}{3!}\bigg(2n\beta_0^2\gamma_0 +3n^2\beta_0\gamma_0^2+n^3\gamma_0^3\bigg)\delta^3\bigg]a_s^3
+\bigg[n\gamma_3\delta\nonumber \\
&+&\frac{1}{2}\bigg(3n\beta_0\gamma_2+2n \beta_1\gamma_1+n\beta_2\gamma_0+2n^2\gamma_0\gamma_2+n^2\gamma_1^2\bigg) \delta^2 +\frac{1}{3!} \bigg(6n\beta_0^2\gamma_1+5n\beta_0\beta_1\gamma_0\nonumber \\
&+&9n^2\beta_0\gamma_0\gamma_1+3 n^2\beta_1\gamma_0^2+3n^3\gamma_0^2\gamma_1\bigg) \delta ^3
+\frac{1}{4!} \bigg(6n\beta_0^3\gamma_0+11n^2\beta_0^2\gamma_0^2+6n^3\beta_0\gamma_0^3\nonumber \\
&+&n^4\gamma_0^4\bigg)\delta ^4 \bigg]a_s^4+\mathcal{O}(a^5_s)\bigg\}. \label{massdis}
\end{eqnarray}

\section{Expression of $\rho_\delta$ up to $a_s^4$-order } \label{B}

\begin{eqnarray}
\rho_\delta(Q^2)&=&r_0 {\overline m}_q^n(\mu_{\delta})\Big\{1+(r_1+n \gamma_0 \delta)a_s(\mu_{\delta}) +\Big[r_2+\beta_0r_1\delta+n(\gamma_1+\gamma_0r_1)\delta+\frac{n}{2}\beta_0\gamma_0\delta^2\nonumber\\
&+&\frac{n^2}{2}\gamma_0^2\delta^2\Big]a^2_s(\mu_{\delta})+\Big[r_3+\beta_1r_1\delta+2\beta_0r_2\delta+\beta_0^2r_1\delta^2+n(\gamma_2+\gamma_0r_2+\gamma_1r_1)\delta\nonumber\\
&+&n\beta_0\Big(\gamma_1+\frac{3\gamma_0r_1}{2}\Big)\delta^2+\frac{n}{2}\beta_1\gamma_0\delta^2+\frac{n}{3}\beta_0^2\gamma_0\delta^3+n^2\Big(\gamma_1\gamma_0+\frac{\gamma_0^2r_1}{2}\Big)
\delta^2+\frac{n^2}{2}\beta_0\gamma_0^2\delta^3\nonumber\\
&+&\frac{n^3}{6}\gamma_0^3\delta^3\Big]a^3_s(\mu_{\delta})+\Big[r_4+(2\beta_1r_2+\beta_2r_1)\delta+3\beta_0r_3\delta+\frac{5}{2}\beta_0\beta_1r_1\delta^2+3\beta_0^2r_2\delta^2\nonumber\\
&+&\beta_0^3r_1\delta^3+n(\gamma_3+\gamma_0r_3+\gamma_1r_2+\gamma_2r_1)\delta+n\beta_0\Big(\frac{3\gamma_2}{2}+\frac{5\gamma_0r_2}{2}+2\gamma_1r_1\Big)\delta^2\nonumber\\
&+&n\Big(\frac{\beta_2\gamma_0}{2}+\beta_1\gamma_1+\frac{3}{2}\beta_1\gamma_0r_1\Big)\delta^2+n\beta_0^2\Big(\gamma_1+\frac{11\gamma_0r_1}{6}\Big)\delta^3+\frac{5n}{6}\beta_0\beta_1\gamma_0\delta^3\nonumber\\
&+&\frac{n}{4}\beta_0^3\gamma_0\delta^4+n^2\Big(\gamma_2\gamma_0+\frac{\gamma_1^2}{2}+\frac{\gamma_0^2r_2}{2}+\gamma_1\gamma_0r_1\Big)\delta^2+n^2\beta_0\Big(\frac{3\gamma_1\gamma_0}{2}+\gamma_0^2r_1\Big)\delta^3\nonumber\\
&+&\frac{n^2}{2}\beta_1\gamma_0^2\delta^3+\frac{11n^2}{24}\beta_0^2\gamma_0^2\delta^4+n^3\Big(\frac{\gamma_1\gamma_0^2}{2}+\frac{\gamma_0^3r_1}{6}\Big)\delta^3+\frac{n^3}{4}\beta_0\gamma_0^3\delta^4+\frac{n^4}{24}\gamma_0^4\delta^4\Big]a^4_s(\mu_{\delta})\nonumber\\
&+&\mathcal{O}[a^5_s(\mu_{\delta})]\Big\}.  \label{rhodeltaB1}
\end{eqnarray}

\section{Expression of $\rho$ up to $a_s^4$-order } \label{C}

\begin{eqnarray}
\rho(Q^2)&=&r_0 {\overline m}_q^n(\mu_r)\Big\{1+\Big({\hat r}_{1,0}+n\gamma_0\ln\frac{\mu_r^2}{Q^2}\Big) a_s(\mu_r) +\Big[{\hat r}_{2,0}+\beta_0{\hat r}_{2,1}+n\gamma_0{\hat r}_{2,1}\nonumber \\
&+&\Big(\beta_0{\hat r}_{1,0}+n\gamma_1+n\gamma_0{\hat r}_{1,0}\Big)\ln\frac{\mu_r^2}{Q^2}+\frac{1}{2}\Big(n\beta_0\gamma_0+n^2\gamma_0^2\Big)\ln^2\frac{\mu_r^2}{Q^2}\Big] a^2_s(\mu_r)\nonumber \\
&+&\Big\{{\hat r}_{3,0}+\beta_1{\hat r}_{2,1}+2\beta_0{\hat r}_{3,1}+\beta_0^2{\hat r}_{3,2}+n(\gamma_0{\hat r}_{3,1}+\gamma_1{\hat r}_{2,1})+\frac{3n}{2}\beta_0\gamma_0{\hat r}_{3,2}\nonumber\\
&+&\frac{n^2}{2}\gamma_0^2{\hat r}_{3,2}+\Big[\beta_1{\hat r}_{1,0}+2\beta_0{\hat r}_{2,0}+2\beta_0^2{\hat r}_{2,1}+n\Big(\gamma_2+\gamma_1{\hat r}_{1,0}+\gamma_0{\hat r}_{2,0}+3\beta_0\gamma_0{\hat r}_{2,1}\nonumber\\
&+&n\gamma_0^2{\hat r}_{2,1}\Big)\Big]\ln\frac{\mu_r^2}{Q^2}+\Big[\beta_0^2{\hat r}_{1,0}+n\Big(\beta_0\gamma_1+\frac{1}{2}\beta_1\gamma_0+\frac{3}{2}\beta_0\gamma_0{\hat r}_{1,0}\Big)+n^2\Big(\gamma_0\gamma_1\nonumber\\
&+&\frac{1}{2}\gamma_0^2{\hat r}_{1,0}\Big)\Big]\ln^2\frac{\mu_r^2}{Q^2}+\frac{1}{6}\Big(2n\beta_0^2\gamma_0+3n^2\beta_0\gamma_0^2+n^3\gamma_0^3\Big)\ln^3\frac{\mu_r^2}{Q^2}\Big\} a^3_s(\mu_r)\nonumber \\
&+&\Big\{{\hat r}_{4,0}+\beta_2{\hat r}_{2,1}+2\beta_1{\hat r}_{3,1}+3\beta_0{\hat r}_{4,1}+\frac{5}{2}\beta_0\beta_1{\hat r}_{3,2}+3\beta_0^2{\hat r}_{4,2}+\beta_0^3{\hat r}_{4,3}\nonumber \\
&+&n\beta_0\Big(2\gamma_1{\hat r}_{3,2}+\frac{5}{2}\gamma_0{\hat r}_{4,2}\Big)+n\Big(\frac{3}{2}\beta_1\gamma_0{\hat r}_{3,2}+\gamma_2{\hat r}_{2,1}+\gamma_1{\hat r}_{3,1}+\gamma_0{\hat r}_{4,1}\Big)\nonumber\\
&+&\frac{11n}{6}\beta_0^2\gamma_0{\hat r}_{4,3}+n^2\Big(\frac{\gamma_0^2{\hat r}_{4,2}}{2}+\gamma_1\gamma_0{\hat r}_{3,2}\Big)+n^2\beta_0\gamma_0^2{\hat r}_{4,3}+\frac{n^3}{6}\gamma_0^3{\hat r}_{4,3}+\Big[\beta_2{\hat r}_{1,0}\nonumber\\
&+&2\beta_1{\hat r}_{2,0}+3\beta_0{\hat r}_{3,0}+6\beta_0^2{\hat r}_{3,1}+3\beta_0^3{\hat r}_{3,2}+5\beta_1\beta_0{\hat r}_{2,1}+n(\gamma_3+\gamma_2{\hat r}_{1,0}+\gamma_1{\hat r}_{2,0}\nonumber\\
&+&\gamma_0{\hat r}_{3,0}+\frac{11}{2}\beta_0^2\gamma_0{\hat r}_{3,2}+4\beta_0\gamma_1{\hat r}_{2,1}+5\beta_0\gamma_0{\hat r}_{3,1}+3\beta_1\gamma_0{\hat r}_{2,1})+n^2(3\beta_0\gamma_0^2{\hat r}_{3,2}\nonumber\\
&+&2\gamma_0\gamma_1{\hat r}_{2,1})+\frac{n^3}{2}\gamma_0^3{\hat r}_{3,2}\Big]\ln\frac{\mu_r^2}{Q^2}+\Big[3\beta_0^2{\hat r}_{2,0}+3\beta_0^3{\hat r}_{2,1}+\frac{5}{2}\beta_0\beta_1{\hat r}_{1,0}+n\Big(\frac{3}{2}\beta_0\gamma_2\nonumber\\
&+&\beta_1\gamma_1+\frac{1}{2}\beta_2\gamma_0+\frac{3}{2}\beta_1\gamma_0{\hat r}_{1,0}+\frac{11}{2}\beta_0^2\gamma_0{\hat r}_{2,1}+2\beta_0\gamma_1{\hat r}_{1,0}+\frac{5}{2}\beta_0\gamma_0{\hat r}_{2,0}\Big)\nonumber\\
&+&n^2\Big(\frac{1}{2}\gamma_1^2+\gamma_0\gamma_2+\gamma_0\gamma_1{\hat r}_{1,0}+\frac{1}{2}\gamma_0^2{\hat r}_{2,0}+3\beta_0\gamma_0^2{\hat r}_{2,1}\Big)+\frac{n^3}{2}\gamma_0^3{\hat r}_{2,1}\Big]\ln^2\frac{\mu_r^2}{Q^2}\nonumber\\
&+&\Big[\beta_0^3{\hat r}_{1,0}+n\Big(\beta_0^2\gamma_1+\frac{11}{6}\beta_0^2\gamma_0{\hat r}_{1,0}+\frac{5}{6}\beta_0\beta_1\gamma_0\Big)+n^2\Big(\beta_0\gamma_0^2{\hat r}_{1,0}+\frac{1}{2}\beta_1\gamma_0^2\nonumber\\
&+&\frac{3}{2}\beta_0\gamma_0\gamma_1\Big)+n^3\Big(\frac{1}{6}\gamma_0^3{\hat r}_{1,0}+\frac{1}{2}\gamma_0^2\gamma_1\Big)\Big]\ln^3\frac{\mu_r^2}{Q^2}+\Big(\frac{n}{4}\beta_0^3\gamma_0+\frac{11n^2}{24}\beta_0^2\gamma_0^2\nonumber\\
&+&\frac{n^3}{4}\beta_0\gamma_0^3+\frac{n^4}{24}\gamma_0^4\Big)\ln^4\frac{\mu_r^2}{Q^2}\Big\} a^4_s(\mu_r)+\mathcal{O}[a^5_s(\mu_r)]\Big\}.
\end{eqnarray}

\section{Relationships between $r_{k}(\mu_r^2/Q^2)(k=3,4)$ and ${\hat r}_{i,j}$ } \label{D}

\begin{eqnarray}
r_3(\mu_r^2/Q^2)&=& {\hat r}_{3,0}+\beta_1{\hat r}_{2,1}+2\beta_0{\hat r}_{3,1}+\beta_0^2{\hat r}_{3,2}+n(\gamma_0{\hat r}_{3,1}+\gamma_1{\hat r}_{2,1})+\frac{3n}{2}\beta_0\gamma_0{\hat r}_{3,2}\nonumber\\
&+&\frac{n^2}{2}\gamma_0^2{\hat r}_{3,2}+\Big[\beta_1{\hat r}_{1,0}+2\beta_0{\hat r}_{2,0}+2\beta_0^2{\hat r}_{2,1}+n\Big(\gamma_2+\gamma_1{\hat r}_{1,0}+\gamma_0{\hat r}_{2,0}+3\beta_0\gamma_0{\hat r}_{2,1}\nonumber\\
&+&n\gamma_0^2{\hat r}_{2,1}\Big)\Big]\ln\frac{\mu_r^2}{Q^2}+\Big[\beta_0^2{\hat r}_{1,0}+n\Big(\beta_0\gamma_1+\frac{1}{2}\beta_1\gamma_0+\frac{3}{2}\beta_0\gamma_0{\hat r}_{1,0}\Big)+n^2\Big(\gamma_0\gamma_1\nonumber\\
&+&\frac{1}{2}\gamma_0^2{\hat r}_{1,0}\Big)\Big]\ln^2\frac{\mu_r^2}{Q^2}+\frac{1}{6}\Big(2n\beta_0^2\gamma_0+3n^2\beta_0\gamma_0^2+n^3\gamma_0^3\Big)\ln^3\frac{\mu_r^2}{Q^2}, \\
r_4(\mu_r^2/Q^2)&=&{\hat r}_{4,0}+\beta_2{\hat r}_{2,1}+2\beta_1{\hat r}_{3,1}+3\beta_0{\hat r}_{4,1}+\frac{5}{2}\beta_0\beta_1{\hat r}_{3,2}+3\beta_0^2{\hat r}_{4,2}+\beta_0^3{\hat r}_{4,3}\nonumber \\
&+&n\beta_0\Big(2\gamma_1{\hat r}_{3,2}+\frac{5}{2}\gamma_0{\hat r}_{4,2}\Big)+n\Big(\frac{3}{2}\beta_1\gamma_0{\hat r}_{3,2}+\gamma_2{\hat r}_{2,1}+\gamma_1{\hat r}_{3,1}+\gamma_0{\hat r}_{4,1}\Big)\nonumber\\
&+&\frac{11n}{6}\beta_0^2\gamma_0{\hat r}_{4,3}+n^2\Big(\frac{\gamma_0^2{\hat r}_{4,2}}{2}+\gamma_1\gamma_0{\hat r}_{3,2}\Big)+n^2\beta_0\gamma_0^2{\hat r}_{4,3}+\frac{n^3}{6}\gamma_0^3{\hat r}_{4,3}+\Big[\beta_2{\hat r}_{1,0}\nonumber\\
&+&2\beta_1{\hat r}_{2,0}+3\beta_0{\hat r}_{3,0}+6\beta_0^2{\hat r}_{3,1}+3\beta_0^3{\hat r}_{3,2}+5\beta_1\beta_0{\hat r}_{2,1}+n(\gamma_3+\gamma_2{\hat r}_{1,0}+\gamma_1{\hat r}_{2,0}\nonumber\\
&+&\gamma_0{\hat r}_{3,0}+\frac{11}{2}\beta_0^2\gamma_0{\hat r}_{3,2}+4\beta_0\gamma_1{\hat r}_{2,1}+5\beta_0\gamma_0{\hat r}_{3,1}+3\beta_1\gamma_0{\hat r}_{2,1})+n^2(3\beta_0\gamma_0^2{\hat r}_{3,2}\nonumber\\
&+&2\gamma_0\gamma_1{\hat r}_{2,1})+\frac{n^3}{2}\gamma_0^3{\hat r}_{3,2}\Big]\ln\frac{\mu_r^2}{Q^2}+\Big[3\beta_0^2{\hat r}_{2,0}+3\beta_0^3{\hat r}_{2,1}+\frac{5}{2}\beta_0\beta_1{\hat r}_{1,0}+n\Big(\frac{3}{2}\beta_0\gamma_2\nonumber\\
&+&\beta_1\gamma_1+\frac{1}{2}\beta_2\gamma_0+\frac{3}{2}\beta_1\gamma_0{\hat r}_{1,0}+\frac{11}{2}\beta_0^2\gamma_0{\hat r}_{2,1}+2\beta_0\gamma_1{\hat r}_{1,0}+\frac{5}{2}\beta_0\gamma_0{\hat r}_{2,0}\Big)\nonumber\\
&+&n^2\Big(\frac{1}{2}\gamma_1^2+\gamma_0\gamma_2+\gamma_0\gamma_1{\hat r}_{1,0}+\frac{1}{2}\gamma_0^2{\hat r}_{2,0}+3\beta_0\gamma_0^2{\hat r}_{2,1}\Big)+\frac{n^3}{2}\gamma_0^3{\hat r}_{2,1}\Big]\ln^2\frac{\mu_r^2}{Q^2}\nonumber\\
&+&\Big[\beta_0^3{\hat r}_{1,0}+n\Big(\beta_0^2\gamma_1+\frac{11}{6}\beta_0^2\gamma_0{\hat r}_{1,0}+\frac{5}{6}\beta_0\beta_1\gamma_0\Big)+n^2\Big(\beta_0\gamma_0^2{\hat r}_{1,0}+\frac{1}{2}\beta_1\gamma_0^2\nonumber\\
&+&\frac{3}{2}\beta_0\gamma_0\gamma_1\Big)+n^3\Big(\frac{1}{6}\gamma_0^3{\hat r}_{1,0}+\frac{1}{2}\gamma_0^2\gamma_1\Big)\Big]\ln^3\frac{\mu_r^2}{Q^2}+\Big(\frac{n}{4}\beta_0^3\gamma_0+\frac{11n^2}{24}\beta_0^2\gamma_0^2\nonumber\\
&+&\frac{n^3}{4}\beta_0\gamma_0^3+\frac{n^4}{24}\gamma_0^4\Big)\ln^4\frac{\mu_r^2}{Q^2}.
\end{eqnarray}

\section{Degeneracy relation } \label{E}

It is possible to infer some degenerate relations in Eq.(\ref{rhodeltaB1}). For example, the coefficients of $\beta_0 a_s^2$, $n\gamma_0 a_s^2$, $\beta_1 a_s^3$, $n\gamma_1 a_s^3$, $\beta_2 a_s^4$, and $n\gamma_2 a_s^4$ are the same, which is $r_1\delta$. By substituting $r_i\delta^j \rightarrow r_{i+j,j}$, the expression for $\rho$ at $\mu_r=Q$ can be rewritten as
\begin{eqnarray}
\rho(Q^2)&=&r_0 {\overline m}_q^n(Q)\Big\{1+{\hat r}_{1,0}a_s(Q) +\left({\hat r}_{2,0}+\beta_0{\hat r}_{2,1}+n\gamma_0{\hat r}_{2,1}\right)a^2_s(Q)+\Big[{\hat r}_{3,0}+\beta_1{\hat r}_{2,1}\nonumber\\
&+&2\beta_0{\hat r}_{3,1}+\beta_0^2{\hat r}_{3,2}+n(\gamma_0{\hat r}_{3,1}+\gamma_1{\hat r}_{2,1})+\frac{3n}{2}\beta_0\gamma_0{\hat r}_{3,2}+\frac{n^2}{2}\gamma_0^2{\hat r}_{3,2}\Big]a^3_s(Q)+\Big[{\hat r}_{4,0}\nonumber\\
&+&\beta_2{\hat r}_{2,1}+2\beta_1{\hat r}_{3,1}+3\beta_0{\hat r}_{4,1}+\frac{5}{2}\beta_0\beta_1{\hat r}_{3,2}+3\beta_0^2{\hat r}_{4,2}+\beta_0^3{\hat r}_{4,3}+n\beta_0\Big(2\gamma_1{\hat r}_{3,2}\nonumber\\
&+&\frac{5}{2}\gamma_0{\hat r}_{4,2}\Big)+n\Big(\frac{3}{2}\beta_1\gamma_0{\hat r}_{3,2}+\gamma_2{\hat r}_{2,1}+\gamma_1{\hat r}_{3,1}+\gamma_0{\hat r}_{4,1}\Big)+\frac{11n}{6}\beta_0^2\gamma_0{\hat r}_{4,3}+n^2\Big(\frac{\gamma_0^2{\hat r}_{4,2}}{2}\nonumber\\
&+&\gamma_1\gamma_0{\hat r}_{3,2}\Big)+n^2\beta_0\gamma_0^2{\hat r}_{4,3}+\frac{n^3}{6}\gamma_0^3{\hat r}_{4,3}\Big]a^4_s(Q)+\mathcal{O}[a^5_s(Q)]\Big\}. \label{rij}
\end{eqnarray}

On the other hand, a pQCD calculation prediction for a physical observable at $\mu_r=Q$ is
\begin{eqnarray}
\rho(Q^2)&=&r_0{\overline m}_q^n(Q)\left[1+\sum^{\infty}_{k=1} \left(\sum^{k-1}_{l=0} c_{k,l} n_f^{l}\right) a_s^k(Q)\right]. \label{cij}
\end{eqnarray}

Comparing the Eq.(\ref{rij}) with Eq.(\ref{cij}) in each order, the coefficients of $n_f$ series can be matched to the $r_{i,j}$ coefficients in a unique way. More explicitly, one can derive the relations between $c_{k,l}$ and $r_{i,j}$ by using the $\beta_i$ and $\gamma_i$ coefficients given in \cite{Baikov:2014qja, Baikov:2016tgj}; e.g., it is easy to find that ${\hat r}_{1,0}=c_{1,0}$. Substituting $\beta_0=11-\frac{2}{3}n_f$ and $\gamma_0=4$ into the $a_s^2$-order coefficient of Eq.(\ref{rij}), we can find $r_{2,0}$ and $r_{2,1}$:
\begin{eqnarray}
c_{2,0}+c_{2,1}n_f={\hat r}_{2,0}+\left(11-\frac{2}{3}n_f\right){\hat r}_{2,1}+4n{\hat r}_{2,1},
\end{eqnarray}
which leads to
\begin{eqnarray}
{\hat r}_{1,0}&=&c_{1,0},\\
{\hat r}_{2,0}&=&c_{2,0}+\left(6n+\frac{33}{2}\right)c_{2,1},~{\hat r}_{2,1}=-\frac{3}{2}c_{2,1}.
\end{eqnarray}
Following the same procedures, one can derive these relations up to any order. In present work, we will use the relations up to fourth order; i.e.,
\begin{eqnarray}
{\hat r}_{3,0}&=&c_{3,0}+\left(3n+\frac{33}{2}\right)c_{3,1}+\left(9n^2+99n+\frac{1089}{4}\right)c_{3,2}-\left(10n^2+11n+\frac{321}{2}\right)c_{2,1},\nonumber\\
{\hat r}_{3,1}&=&-\frac{3}{4}c_{3,1}-\frac{27n+99}{4}c_{3,2}+\frac{10n+57}{4}c_{2,1},~{\hat r}_{3,2}=\frac{9}{4}c_{3,2}, \\
{\hat r}_{4,0}&=&c_{4,0}+\Big(2n+\frac{33}{2}\Big)c_{4,1}+\Big(4n^2+66n+\frac{1089}{4}\Big)c_{4,2}+\Big(8n^3+198n^2+\frac{3267n}{2}\nonumber\\
&&+\frac{35937}{8}\Big)c_{4,3}-\Big(\frac{50n^3}{3}+185n^2+\frac{2595n}{2}+\frac{10593}{2}\Big)c_{3,2}-\Big(\frac{10n^2}{3}+15n+\frac{321}{2}\Big)c_{3,1}\nonumber\\
&&+\Big(\frac{20n^3}{27}-160\zeta_3n^2-\frac{2411n^2}{9}-\frac{860n}{3}-1320\zeta_3n+\frac{11675}{16}\Big)c_{2,1},\\
{\hat r}_{4,1}&=&-\frac{1}{2}c_{4,1}-\frac{1}{2}\left(5n+33\right)c_{4,2}-\left(\frac{19n^2}{2}+\frac{495n}{4}+\frac{3267}{8}\right)c_{4,3}\nonumber+\left(\frac{5n}{6}+\frac{19}{2}\right)c_{3,1}\\
&&+\left(\frac{85n^2}{6}+\frac{401n}{4}+\frac{4113}{8}\right)c_{3,2}+\left(\frac{100n^2}{27}+40\zeta_3n+\frac{2467n}{72}-\frac{479}{4}\right)c_{2,1},\\
{\hat r}_{4,2}&=&\frac{3}{4}c_{4,2}+\left(\frac{33n}{4}+\frac{297}{8}\right)c_{4,3}-\left(5n+\frac{285}{8}\right)c_{3,2}+\left(\frac{325}{48}-\frac{35n}{18}\right)c_{2,1},\\
{\hat r}_{4,3}&=&-\frac{27}{8}c_{4,3}.
\end{eqnarray}
It should be noted that one must treat the $n_f$ terms unrelated to renormalization of the running coupling and running mass as part of the conformal coefficient; e.g., the $n_f$ terms coming from the light-by-light diagrams in QCD belongs to the conformal series. The contributions of light-by-light diagrams is always given separately, since the light-by-light diagrams can be easily distinguished from the other Feynman diagrams.

\section{ $r_{\rm i, NonConf.}(i=3,4)$ } \label{F}

\begin{eqnarray}
r_{\rm 3, NonConf.}(Q_*)&=& \beta_1{\hat r}_{2,1}+2\beta_0{\hat r}_{3,1}+\beta_0^2{\hat r}_{3,2}+n(\gamma_0{\hat r}_{3,1}+\gamma_1{\hat r}_{2,1})+\frac{3n}{2}\beta_0\gamma_0{\hat r}_{3,2}\nonumber\\
&+&\frac{n^2}{2}\gamma_0^2{\hat r}_{3,2}+\Big[\beta_1{\hat r}_{1,0}+2\beta_0{\hat r}_{2,0}+2\beta_0^2{\hat r}_{2,1}+n\Big(\gamma_2+\gamma_1{\hat r}_{1,0}+\gamma_0{\hat r}_{2,0}+3\beta_0\gamma_0{\hat r}_{2,1}\nonumber\\
&+&n\gamma_0^2{\hat r}_{2,1}\Big)\Big]\ln\frac{Q_*^2}{Q^2}+\Big[\beta_0^2{\hat r}_{1,0}+n\Big(\beta_0\gamma_1+\frac{1}{2}\beta_1\gamma_0+\frac{3}{2}\beta_0\gamma_0{\hat r}_{1,0}\Big)+n^2\Big(\gamma_0\gamma_1\nonumber\\
&+&\frac{1}{2}\gamma_0^2{\hat r}_{1,0}\Big)\Big]\ln^2\frac{Q_*^2}{Q^2}+\frac{1}{6}\Big(2n\beta_0^2\gamma_0+3n^2\beta_0\gamma_0^2+n^3\gamma_0^3\Big)\ln^3\frac{Q_*^2}{Q^2}, \\
r_{\rm 4, NonConf.}(Q_*)&=& \beta_2{\hat r}_{2,1}+2\beta_1{\hat r}_{3,1}+3\beta_0{\hat r}_{4,1}+\frac{5}{2}\beta_0\beta_1{\hat r}_{3,2}+3\beta_0^2{\hat r}_{4,2}+\beta_0^3{\hat r}_{4,3}\nonumber \\
&+&n\beta_0\Big(2\gamma_1{\hat r}_{3,2}+\frac{5}{2}\gamma_0{\hat r}_{4,2}\Big)+n\Big(\frac{3}{2}\beta_1\gamma_0{\hat r}_{3,2}+\gamma_2{\hat r}_{2,1}+\gamma_1{\hat r}_{3,1}+\gamma_0{\hat r}_{4,1}\Big)\nonumber\\
&+&\frac{11n}{6}\beta_0^2\gamma_0{\hat r}_{4,3}+n^2\Big(\frac{\gamma_0^2{\hat r}_{4,2}}{2}+\gamma_1\gamma_0{\hat r}_{3,2}\Big)+n^2\beta_0\gamma_0^2{\hat r}_{4,3}+\frac{n^3}{6}\gamma_0^3{\hat r}_{4,3}+\Big[\beta_2{\hat r}_{1,0}\nonumber\\
&+&2\beta_1{\hat r}_{2,0}+3\beta_0{\hat r}_{3,0}+6\beta_0^2{\hat r}_{3,1}+3\beta_0^3{\hat r}_{3,2}+5\beta_1\beta_0{\hat r}_{2,1}+n(\gamma_3+\gamma_2{\hat r}_{1,0}+\gamma_1{\hat r}_{2,0}\nonumber\\
&+&\gamma_0{\hat r}_{3,0}+\frac{11}{2}\beta_0^2\gamma_0{\hat r}_{3,2}+4\beta_0\gamma_1{\hat r}_{2,1}+5\beta_0\gamma_0{\hat r}_{3,1}+3\beta_1\gamma_0{\hat r}_{2,1})+n^2(3\beta_0\gamma_0^2{\hat r}_{3,2}\nonumber\\
&+&2\gamma_0\gamma_1{\hat r}_{2,1})+\frac{n^3}{2}\gamma_0^3{\hat r}_{3,2}\Big]\ln\frac{Q_*^2}{Q^2}+\Big[3\beta_0^2{\hat r}_{2,0}+3\beta_0^3{\hat r}_{2,1}+\frac{5}{2}\beta_0\beta_1{\hat r}_{1,0}+n\Big(\frac{3}{2}\beta_0\gamma_2\nonumber\\
&+&\beta_1\gamma_1+\frac{1}{2}\beta_2\gamma_0+\frac{3}{2}\beta_1\gamma_0{\hat r}_{1,0}+\frac{11}{2}\beta_0^2\gamma_0{\hat r}_{2,1}+2\beta_0\gamma_1{\hat r}_{1,0}+\frac{5}{2}\beta_0\gamma_0{\hat r}_{2,0}\Big)\nonumber\\
&+&n^2\Big(\frac{1}{2}\gamma_1^2+\gamma_0\gamma_2+\gamma_0\gamma_1{\hat r}_{1,0}+\frac{1}{2}\gamma_0^2{\hat r}_{2,0}+3\beta_0\gamma_0^2{\hat r}_{2,1}\Big)+\frac{n^3}{2}\gamma_0^3{\hat r}_{2,1}\Big]\ln^2\frac{Q_*^2}{Q^2}\nonumber\\
&+&\Big[\beta_0^3{\hat r}_{1,0}+n\Big(\beta_0^2\gamma_1+\frac{11}{6}\beta_0^2\gamma_0{\hat r}_{1,0}+\frac{5}{6}\beta_0\beta_1\gamma_0\Big)+n^2\Big(\beta_0\gamma_0^2{\hat r}_{1,0}+\frac{1}{2}\beta_1\gamma_0^2\nonumber\\
&+&\frac{3}{2}\beta_0\gamma_0\gamma_1\Big)+n^3\Big(\frac{1}{6}\gamma_0^3{\hat r}_{1,0}+\frac{1}{2}\gamma_0^2\gamma_1\Big)\Big]\ln^3\frac{Q_*^2}{Q^2}+\Big(\frac{n}{4}\beta_0^3\gamma_0+\frac{11n^2}{24}\beta_0^2\gamma_0^2\nonumber\\
&+&\frac{n^3}{4}\beta_0\gamma_0^3+\frac{n^4}{24}\gamma_0^4\Big)\ln^4\frac{Q_*^2}{Q^2}.
\end{eqnarray}

\section{ $S_3$} \label{G}

\begin{eqnarray}
S_3&=&{\hat r}_{1,0}{\hat r}_{3,1}-{\hat r}_{1,0}^2{\hat r}_{2,1}+{\hat r}_{2,0}{\hat r}_{2,1}-{\hat r}_{4,1}+\frac{n\gamma_0}{2}({\hat r}_{2,1}^2+{\hat r}_{1,0}{\hat r}_{3,2}-{\hat r}_{4,2})-\frac{n\gamma_1{\hat r}_{3,2}}{2}-\frac{n^2\gamma_0^2{\hat r}_{4,3}}{6}\nonumber \\
&-&\frac{\beta_2{\hat r}_{2,1}}{n\gamma_0}+\beta_1\Big(\frac{\gamma_1{\hat r}_{2,1}}{n\gamma_0^2}+\frac{2{\hat r}_{1,0}{\hat r}_{2,1}-2{\hat r}_{3,1}}{n\gamma_0}-\frac{3{\hat r}_{3,2}}{2}\Big)+\beta_0\beta_1\Big(\frac{2{\hat r}_{1,0}{\hat r}_{2,1}}{n^2\gamma_0^2}-\frac{5{\hat r}_{3,2}}{2n\gamma_0}\Big)\nonumber \\
&+&\beta_0\Bigg(\frac{5{\hat r}_{2,1}^2}{2}-n\gamma_0{\hat r}_{4,3}-\frac{5{\hat r}_{4,2}}{2}+2{\hat r}_{1,0}{\hat r}_{3,2}-\frac{\gamma_1^2{\hat r}_{2,1}}{n\gamma_0^3}+\frac{\gamma_2{\hat r}_{2,1}+2\gamma_1{\hat r}_{3,1}-2\gamma_1{\hat r}_{1,0}{\hat r}_{2,1}}{n\gamma_0^2}\nonumber \\
&-&\frac{n\gamma_1{\hat r}_{3,2}+6{\hat r}_{2,1}{\hat r}_{1,0}^2-6{\hat r}_{3,1}{\hat r}_{1,0}-6{\hat r}_{2,0}{\hat r}_{2,1}+6{\hat r}_{4,1}}{2n\gamma_0}\Bigg)+\beta_0^2\Bigg(\frac{7{\hat r}_{2,1}^2+5{\hat r}_{1,0}{\hat r}_{3,2}-6{\hat r}_{4,2}}{2n\gamma_0}\nonumber\\
&+&\frac{n\gamma_1{\hat r}_{3,2}-3{\hat r}_{2,1}{\hat r}_{1,0}^2+2{\hat r}_{3,1}{\hat r}_{1,0}+2{\hat r}_{2,0}{\hat r}_{2,1}}{n^2\gamma_0^2}-\frac{2\gamma_1{\hat r}_{1,0}{\hat r}_{2,1}}{n^3\gamma_0^3}-\frac{11{\hat r}_{4,3}}{6}\Bigg)\nonumber\\
&+&\beta_0^3\Big(\frac{3{\hat r}_{2,1}^2+2{\hat r}_{1,0}{\hat r}_{3,2}}{2n^2\gamma_0^2}-\frac{{\hat r}_{2,1}{\hat r}_{1,0}^2}{n^3\gamma_0^3}-\frac{{\hat r}_{4,3}}{n\gamma_0}\Big).
\end{eqnarray}

\end{widetext}

\end{document}